\documentclass[reprint,amsmath,amssymb,aps,twocolumn]{revtex4-1}
\usepackage{graphicx}
\usepackage{dcolumn}
\usepackage{bm}
\usepackage{braket}
\usepackage{hyperref}

\begin{document}
\title{A magnetic tight-binding model: correlations in ferromagnetic transition metals}% Force line breaks with \\
\author{Jacques R. Eone II\\} 
% \altaffiliation[Also at ]{Physics Department, Strasbourg University.}%Lines break automatically or can be forced with \\ssss
% \email{jacques.eone@ipcms.unistra.fr}
\affiliation{%
 \\ Department of Physics, University of Strasbourg,  Strasbourg, France
}%

\date{2018} 

\begin{abstract}
Correlations derived through single-particle approximations of the many-body problem frequently result in erroneously inflated or diminished physical properties. In the context of transition metals, the impact of correlations can be assessed by analyzing the effect of the delocalized $sp$-band on the $d$-band. The tight-binding approach is studied in the $d$-band approximation, considering and excluding the influence of $sp-d$ hybridization.  The impact of the delocalized $sp$-band induces a correction to the onsite Coulomb parameter. This formalism enables an accurate description of the ferromagnetism within the tight-binding approximation. The onsite Coulomb corrections, which were calculated in accordance with first-principles results, are 1.3 eV for bcc iron, 1.5 eV for fcc cobalt, and 2.1 eV for fcc nickel.
\end{abstract} 

\pacs{Valid PACS appear here}
\maketitle

\section{\label{sec:level1}Introduction}
Density functional theory (DFT) is a highly effective method for calculating the electronic structure of solids \cite{cite1}. However, the precision of this approach is determined by the exchange-correlation functional employed. Consequently, the DFT results may exhibit under- or overestimation of physical properties depending on the selected functional, along with an erroneous estimation of the bandgap in semiconductors or insulators. A variety of methodologies have been employed to address these limitations, including the GW approximation \cite{cite2}, hybrid functionals \cite{cite3}, and Hubbard corrections \cite{cite4}. The objective of this study is not to examine these corrections, but rather to investigate the physical process induced by Coulomb correlations and its relationship to local magnetism in a semi-empirical model using the tight-binding approximation. The magnetism can be integrated in the tight-binding approximation, limited to the $d$-band using the Stoner model from a local Hubbard Hamiltonian \cite{cite5}. This implementation results in an overestimation of the Coulomb $U$ parameter. However, it provides an accurate description of the energies \cite{cite6,cite7}. This formalism employs a $d$-band with the effect of the $sp-d$ hybridization. An alternative approach involves a model of magnetism with the onsite Coulomb parameter that aligns with photoemission experiments. However, this model's description of energies is not in agreement with \textit{ab initio} calculations \cite{cite8}. The approach is constructed with a localized $d$-band, excluding the $sp-d$ hybridization effect.
These contradictory outcomes are attributable to the approximate characterization of the Coulomb correlations. In the context of solids, the interactions including Coulomb repulsion are a competition to reduce the total energy of the system. The $d$-orbitals interact with the environment, thereby reducing the Coulomb repulsion and kinetic energies, and consequently generating a cohesive energy. However, the $d$-orbitals exhibit a degree of localization in comparison to the $s$ orbital of transition metals. The metallicity of the ground state involving $d$ electrons is also attributable to the delocalized $s$ state. The coupling between localized ($d$) and delocalized states ($sp$) gives rise to the overall metallic behavior exhibited by transition metals. This metallicity can be described by a single-particle density functional theory. In the context of strongly correlated systems, another coupling emerges, resulting in a state that is no longer metallic. The present study analyzes the impact of the $sp$-band on the $d$-band. The properties of transition metals are determined by the impact of the $s$ electron, which spans the $s$, $p$, and $d$ bands. In this study, the effect of delocalized states on the $d$-band is analyzed using the onsite Coulomb interaction, which is related to the spin magnetic moment of ferromagnetic metals.
The objective of this study is to elucidate the correlation effects of the $sp-d$ coupling within the framework of a simplified magnetic tight-binding approximation, encompassing both the contribution of the $sp$-band and its absence.

\section{\label{sec:level2}Methodology}
In a mean field approximation, the local Hubbard Hamiltonian can be expressed as follows: 
%The tight-binding approximation in a mean field theory (Hartree-Fock) adopts a local form of the Hubbard model in the \textit{d}-band.
\begin{equation}
H_i=- t \sum_{j,\sigma} \left( c_{i\sigma}^\dagger  c_{j\sigma} + h.c. \right)+U_d\sum_\lambda  n_{\lambda\uparrow} n_{\lambda\downarrow},
\label{eq:eq1}
\end{equation}

where $i$ denotes an atomic site, while $j$ indicates the neighboring sites. This Hamiltonian is employed to determine the local electronic structure, wherein the parameter $\lambda$ denotes an orbital, not an atomic site. The second term of the Hamiltonians is expressed as a summation over five d-orbitals.  The effective Coulomb repulsion on a $d$ orbital is denoted by $U_d$. $t$ is the hopping parameter. In the tight-binding approximation, the atomic energies $\epsilon_d$ of the $d$ electrons are supplementary parameters. These can be decomposed into an atomic contribution and a perturbation $\alpha$: 
\begin{equation}
\epsilon_d =\underbrace{\int \psi_\lambda^*(\bm{r}) H^\text{at}\psi_\lambda(\bm{r}) d^3 r}_{\text{atomic}}+\underbrace{\int \psi_\lambda^*(\bm{r})\Delta U(\bm{r})\psi_\lambda(\bm{r}) d^3 r}_{\alpha \text{ : pertubation}},
\end{equation}
where $\psi_\lambda$ represents the atomic basis, while $H^\text{at}$ denotes the atomic Hamiltonian. $\alpha$ is  the first-order correction to the atomic energy in a perturbation $\Delta U(r)$ due to the interaction with neighboring atoms. Due to the weak overlap of the $d$ orbitals, this correction is frequently disregarded. This correction suggests that any variation in the atomic potential, caused by a perturbation, results in a shift of the atomic levels.  On the other hand, each shift in atomic energies can be attributed to the appearance of a  perturbation. This assertion can be employed to formulate a self-consistency correction for systems with a surface.  The interaction with the neighbors creates a band of width $W$; this width is defined as the range of possible electron energies within the solid. The bandwidth is proportional to the hopping integral composed of an atomic orbital $\lambda$ of the site $i$ and an atomic orbital $\nu$ of a neighboring atom:
\begin{equation}
W \propto -\int \psi_\lambda^*(\bm{r})\Delta U(r)\psi_\nu(\bm{r}-\bm{R}_j) d^3 r
\end{equation}
The local Hubbard Hamiltonian of Eq. (\ref{eq:eq1}) provides the effective Coulomb repulsion $U_d$  which is restricted to the $d$ orbitals. Assuming that the hopping parameters are invariant, the effect of the $sp$-band can be captured by implementing a correction $U_d^{sp}$ on the effective Coulomb repulsion:
\begin{eqnarray*}
H_i=- t \sum_{j,\sigma} \left( c_{i\sigma}^\dagger  c_{j\sigma} + h.c. \right)+U_d \sum_\lambda n_{\lambda\uparrow} n_{\lambda\downarrow} \\ + U_d^{sp} \sum_\lambda n_{\lambda\uparrow} n_{\lambda\downarrow} 
\end{eqnarray*}

In most cases, the presence of a supplementary onsite Coulomb repulsion results in a localization. However, this local approximation assumes that the supplementary repulsion leads to a more delocalized ground state with the $sp-d$ coupling compared to the ground state in the absence of the $sp-d$ hybridization. \\

The Coulomb repulsion parameter $U_d$ containing the correlations can be related to the $d$-band ferromagnetism. According to the  Eq. (\ref{eq:eq1}), by considering $n_0=5$  $d$ orbitals containing $n_d$ electrons,  the local spin magnetic moment $\mu$ and the number of electrons $n_d$ in the $d$-band are given by:  $\mu=n_0 \left\langle n_\uparrow  -n_\downarrow \right\rangle  \text{ and } n_d=n_0 \left\langle n_\uparrow  + n_\downarrow \right\rangle$. The  Coulomb contribution of the Hamiltonian can be written as (a similar result in a lattice can be found in the literature \cite{cite9}): 
\begin{eqnarray*}
U_d\sum_\lambda n_{\lambda\uparrow} n_{\lambda\downarrow} \approx U_d \sum_\lambda n_{\lambda\uparrow} \left\langle n_{\downarrow} \right\rangle + n_{\lambda\downarrow} \left\langle n_{\uparrow} \right\rangle -\left\langle n_{\uparrow} \right\rangle \left\langle n_{\downarrow} \right\rangle \\
=U_d\sum_{k,\sigma} n_{k\sigma} \left\langle n_{-\sigma} \right\rangle -n_0U_d \left\langle n_{\uparrow} \right\rangle \left\langle n_{\downarrow} \right\rangle \\
\end{eqnarray*}
\begin{align*}
&=\frac{U_d}{2n_0} \sum_{k\sigma} \left( n_d-\sigma \mu\right) c_{k\sigma}^\dagger c_{k\sigma} - n_0 U_d\frac{1}{4n_0^2}(n_d-\mu)(n_d+\mu)\\
&=\frac{U_d}{n_0} \sum_{k\sigma} \left(\frac{n_d}{2}- \frac{\sigma }{2}\mu\right) c_{k\sigma}^\dagger c_{k\sigma} - \frac{U_d}{n_0}\left(\frac{n_d^2}{4}-\frac{\mu^2}{4}\right) 
\end{align*}
If the stoner criterion is fulfilled, the band structure  $\epsilon_{k \sigma} =\epsilon_k + \frac{n_dU_d}{2n_0} -\frac{\sigma }{2}\frac{U_d\mu}{n_0} $  is therefore dependent on the spin  $\sigma$ and the bands per spin are shifted by the exchange splitting: 
\begin{equation}
\Delta \epsilon =  \frac{U_d\mu}{n_0} = I_{d}\mu,
\label{eq:mu_scf}
\end{equation}
where $I_{d}$ is the Stoner parameter. The Eq. (\ref{eq:eq1})  can be written as:  
\begin{equation}
H_i=\sum_{k\sigma} \left(\epsilon_k + \frac{n_d U_d}{2n_0} -\frac{\sigma }{2} \frac{U_d\mu}{n_0}\right) c_{k\sigma}^\dagger c_{k\sigma} - \frac{U_d}{n_0}\left(\frac{n_d^2}{4}-\frac{\mu^2}{4}\right) 
\label{eq:tb_etot}
\end{equation}
The Stoner relation of Eq. (\ref{eq:mu_scf}) can also be deduced by assuming that the exchange splitting $\Delta\epsilon$ is a potential energy due to the reduction of the Coulomb repulsion $U_d$ of $\mu$ unpaired electrons in $n_0$ $d$-orbitals.  Using  Eq. (\ref{eq:tb_etot}), the variation of the binding energy $E_{b}$ from a nonmagnetic state to a ferromagnetic state can be deduced:
\begin{align*}
 \Delta E^{mag.}&= E_b^\text{ferro.}-E_b^\text{nonmag.}\\
				&= \Delta E_b -\frac{1}{4n_0}U_d\mu^2  (\text{delocalized})\\
				&= \Delta E_b -\frac{1}{4n_0}U_d^\text{loc.}\mu^2  -  \frac{1}{4n_0}U_d^{sp}\mu^2 (\text{localized}),
\label{eq:e_mag}				
\end{align*}
where $\Delta E_b = E_b^\uparrow + E_b^\downarrow - E_b^\text{nonmag.}$. Given the stability of the magnetic state in a ferromagnetic metal, the energy change $ \Delta E^\text{mag.}$ should be always negative. The term $U_d^{sp}$  is employed in the localized $d$-band approximation where the effects of the $sp$-band are neglected, to correct the magnetic ground state. In the delocalized case, the effective Coulomb repulsion $U_d$ already incorporates all the corrections. In this work, the quantity $U_d^{sp}$ is obtained through the calculation of the local density of states (LDOS) $n_d(E)$ using the tight-binding approximation.
 
The tight-binding approximation enables the calculation of a localized $d$-LDOS $n_d^\text{loc.}(E)$ by diagonalizing a $5\times 5$ Hamiltonian matrix consisting of three $d$ hopping parameters ($dd\sigma, dd\pi, dd\delta$). In this calculation, the hopping parameters of the $s$ and $p$ orbitals are neglected. A delocalized $d$-LDOS $n_d^\text{deloc.}(E)$ is obtained through the diagonalization of a $9\times 9$ matrix containing all the orbitals. The correction $U_d^{sp}$ is determined approximately by calculating the difference between the band energy of a localized and a delocalized $d$-LDOS: 
\begin{equation}
U_d^{sp}\approx \int_{-\infty}^{E_F} E n_d^\text{loc.}(E)dE -\int_{-\infty}^{E_F} E n_d^\text{deloc.}(E) dE
\label{eq:corr}
\end{equation}
The parameter $U_d^{sp}$ incorporates the screening effects and the impact of the $sp$-band on the $d$-band. In the event that this parameter is found to be substantial, the localized approach is no longer capable of providing a comprehensive description of the transition metal electronic structure.  As illustrated by Eq. (\ref{eq:corr}), the quantity $U_d^{sp}$ is significantly impacted by the charge in the $d$-band. Therefore, the calculated $U_d^{sp}$ parameters are the values corresponding to the first-principles $ \Delta E^\text{mag.}$ energy change  using the localized approximation. \\

The quantity $U_d$ can be obtained by using the following relation \cite{cite12,cite13}: 
\begin{equation}
U_d = E(n_d+1) + E(n_d-1) - 2 E(n_d)  
\label{eq:curie}
\end{equation}
$E$ is used to denote a total energy that, unfortunately, cannot be calculated with the local density of states within the tight-binding approximation. In this study, $U_d$ is determined using the density of states and the experimental spin magnetic moment.

\section{\label{sec:level3}Results}

\begin{figure}[!h]
	\includegraphics[width=0.4\textwidth]{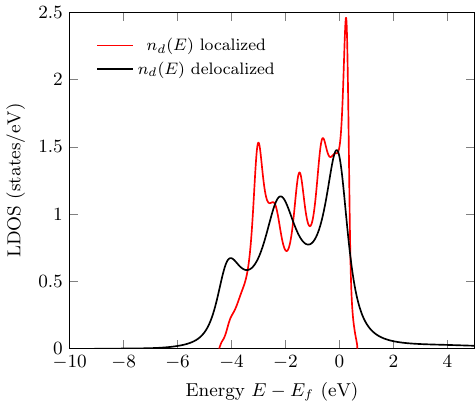}% Here is how to import EPS art
	\caption{\label{fig:fig1} Comparison of the LDOS of a localized  $d$-band and a delocalized $d$-band.}
\end{figure}

The hopping parameters of D. A. Papaconstantopoulos \cite{cite14} are used in this study in order to calculate the localized and delocalized $d$-LDOS. As illustrated in Fig. \ref{fig:fig1}, the localized $d$-LDOS exhibits a reduced bandwidth in comparison to the delocalized $d$-LDOS.

A reduced bandwidth indicates that the electrons are localized, thereby contradicting the metallic nature typically associated with transition metals. In this localized state, the $s$ and $p$ electrons  do not interact with the $d$-band. This localized $d$-LDOS is analogous to that of strongly correlated systems, where the $d$ orbitals are markedly localized. The description of a transition metal employing a localized $d$-LDOS results in the determination of binding or surface energies that are not accurate. Consequently, the implementation of corrections is necessary.  The delocalized $d$-LDOS encompasses the effects of the $sp$-band on the $d$-band, constituting the ground state of a transition metal.\\

The $U_d$ parameter is determined through the fit of the spin magnetic moment, which is obtained using two different equations \cite{cite8}. The first equation is derived from the shift of rigid LDOS, employing distinct values of $\Delta \epsilon$, while maintaining the same charge in the $d$-band. 
\begin{equation}
\mu (\Delta \epsilon)=n_{d\uparrow}(E) - n_{d\downarrow}(E)  = n_d(E -  \frac{\Delta \epsilon}{2})  +  n_d(E +  \frac{\Delta \epsilon}{2}) 
\end{equation}
The second equation is derived from the Stoner relation in Eq. (\ref{eq:mu_scf}):
\begin{equation}
 \mu (\Delta \epsilon)= \frac{5}{U_d} \Delta \epsilon
\end{equation}

The calculation performed using a localized $d$-LDOS yielded a value of 3.04 eV for fcc Co. This value is obtained by shifting the LDOS by an exchange splitting of $0.98$ eV to obtain a spin magnetic moment of approximately  $\mu=1.61 \text{ }\mu_B$. This result is consistent with the photoemission results, which indicate that $\Delta \epsilon=1.0$ eV \cite{cite15,cite16,cite17}. In a similar manner, the magnitude of $U_d$ for both bcc Fe and fcc Ni are obtained to be $2.93$ eV and $2.21$ eV, respectively. A result obtained by shifting the $d$-LDOS with respective $\Delta \epsilon$ of $1.29$ eV and $0.27$ eV. The obtained values are consistent with photoemission exchange splittings \cite{cite18,cite19} for the experimental spin magnetic moment of 2.2 $\mu_B$ in bcc Fe  and 0.61 $\mu_B$ in fcc Ni. These quantities are summarized in Table \ref{tab:table1}, where $\Delta E^\text{mag.}= \Delta E_b -\frac{1}{4n_0}U_d^\text{loc.}\mu^2$ is calculated without correction.
\begin{table}[h!]
\caption{\label{tab:table1} Calculated values of $U_d$, $\Delta \epsilon $ and $\Delta E^\text{mag.}$  for a localized $d$-band.}
\begin{ruledtabular}
\begin{tabular}{lccc}
                                    & BCC Fe  & FCC Co & FCC Ni \\\hline
$U_d^\text{loc.}$ [eV]         &  2.93 &  3.04 & 2.21   \\
$\mu$ [$\mu_B$]                   &  2.20 & 1.61 & 0.61  \\
$\Delta \epsilon$ [eV]   &  1.29 & 0.98 & 0.27   \\
$I_d$ [eV]                     &  0.98 & 0.61 & 0.44   \\
$\Delta E^\text{mag.}$ [eV]    & -0.13  & 0.02  & -0.004 \\
\end{tabular}
\end{ruledtabular}
\end{table}

As would be expected, the localized approach does not accurately describe the $d$-band magnetism, resulting in underestimated values and energies that are not in agreement with the stability of the ferromagnetic state. This is despite the approach's alignment with the results of photoemission. It has been observed that the value of the Coulomb parameter $U_d^\text{loc.}$ is expected to increase with the filling of the $d$-band \cite{cite11}. However, this effect does not appear to be observed in the present context.  It has been established that a total energy calculated with this localized approximation leads to surface energies that are half the value obtained with first-principles calculations. The $U_d^{sp}$ parameters correcting the magnitude of $\Delta E^\text{mag.}$ for the localized $d$-band approximation can be calculated. The following values were obtained:  1.3 eV for  bcc Fe, 1.5 eV for fcc Co and 2.1 eV for  fcc Ni. These values are analogous to those reported previously in a separate study on the correction of band structures \cite{cite10}. \\

In the case of a delocalized $d$-band, the exchange splittings for  bcc Fe, fcc Co and fcc Ni  are 2.23 eV, 1.94 eV and 0.82 eV respectively (comparable to the results of another calculation \cite{cite20,cite21,cite22}). The $U_d$ parameters for the same metals are calculated to be 4.98 eV, 6.02 eV and 6.72 eV respectively. These quantities are summarized in the Table \ref{tab:table2}. 
 
\begin{table}[h!]
\caption{\label{tab:table2} Calculated values of $U_d$, $\Delta \epsilon $ and $\Delta E^\text{mag.}$ for a delocalized $d$-band.}
\begin{ruledtabular}
\begin{tabular}{lccc}
                                              & BCC Fe  & FCC Co   & FCC Ni  \\\hline
$U^\text{deloc.}$ [eV]                   &  4.98    & 6.02     & 6.72   \\
$\mu$ [$\mu_B$]                            &  2.24     & 1.61      & 0.61  \\
$\Delta \epsilon$ [eV]            &  2.23    & 1.94      & 0.82   \\
$I_d$ [eV]                              &  0.99   &  1.20      & 1.34  \\
$\Delta E^\text{mag.} $ [eV]          &  -0.40  & -0.22    & -0.035 \\
\end{tabular}
\end{ruledtabular}
\end{table}
The calculated values are consistent with the ferromagnetic state, in which the energy difference $\Delta E^\text{mag.}$ is negative for the three ferromagnetic metals under consideration. Contrary to the localized approximation, a consistent value of $\Delta E^\text{mag.}$ is obtained, which is in agreement with first-principles calculations. However, the Coulomb $U_d$ parameter appears to be overestimated.   \\

%This delocalized approximation on the other hand does not correctly describe the case where the $d$-band is localized as in transition metal oxides. In this case the $s$ and $p$ electrons are involved in  covalent bonds and the $d$-band is less impacted by them. The localized approximation then describes the properties of the $d$-band. A delocalized approach as in a density functional theory calculation with an uncorrected exchange-correlation  functional unfortunately leads to describe them as metals. The gap $E_g$ in this case connects the two approaches as the Coulomb barrier to cross to delocalize  a $d$ electron define by :
%\begin{equation}
%E_g = U_d^{deloc.} -  U_d^{loc.}
%\end{equation}
%This relation used simply in our case can not be generalized to all transition metals. The calculated gaps  $E_g$  for the NiO, FeO et CoO are  respectively 2.05 eV, 2.98  eV et 4.51 eV (in agreement with values obtained in the work of  ~\citealt{cite23} and ~\citealt{cite24}).  \\ \\

\section{\label{sec:level4}Conclusion}
The interactions between electrons in a solid are fundamental to solid-state physics. The means to describe them remain approximate until the electronic correlations are accurately implemented. As demonstrated in this study, the correlations facilitate the connection between two methodologies employed to describe the $d$-band of a transition metal. These two approaches, termed "localized" and "delocalized", which appear to be disparate, are, in fact, closely related. While the localized model appears to be a straightforward approach for describing the $d$-band of a transition metal, it is inadequate for fully capturing the energies of these materials. In instances where the impact of the $sp$-band on the $d$-band is significant, the localized model becomes inadequate. Consequently, a delocalized approach is required to accurately depict the metallic behavior. Despite the implementation of corrections, such as the one applied to the magnetic energy for the localized model, these adjustments pose significant challenges in their application to all physical scenarios. In this study, the term "delocalized" is employed to denote a state of reduced localization. The $d$-band consistently exhibits a degree of localization compared to the $s$- and $p$-bands.   The tight-binding approximation is a widely utilized and a valuable method for elucidating the properties of complex materials by parameterizing the effects of electronic correlations. The semi-empirical magnetism presented here can be extended to the study of magnetic surfaces and nanoparticles, as well as to determine the properties of transition metal alloys.

\bibliography{article}

\end{document}